# Spatially incoherent modulational instability in a non local medium

**Marco Peccianti, Claudio Conti, Emiliano Alberici and Gaetano Assanto**

N*OO*EL - *Nonlinear Optics and Optoelectronics Laboratory*

*National Institute for the Physics of Matter - University "Roma Tre*

*Via della Vasca Navale 84, 00146, Rome, Italy*

*Te.l +39-0655177028; Fax +39-065579078, E-mail assanto@ele.uniroma3.it*

We investigate one-dimensional transverse modulational instability in a non local medium excited with a spatially incoherent source. Employing undoped nematic liquid crystals in a planar pre-tilted configuration, we investigate the role of the spectral broadening induced by incoherence in conjunction with the spatially non local molecular reorientation. The phenomenon is modeled using the Wigner transform.

Modulational instability (MI) in time and space is generally related to the instability of flat wavefront solutions in the presence of a nonlinear response. Optical transverse MI is usually thought as a precursor to spatial solitons [1] and it has been previously investigated in parametric crystals, [2-5] photorefractives, [6-11] semiconductors, [12] and liquid crystals. [13] Noise perturbations in an intense enough plane wave produce a transverse pattern due to the nonlinear response, which can evolve into filaments. The growth of transverse spectral components drags energy from the input plane wave, with a resulting selective gain peaked at a specific periodicity for a given excitation. [1,14-15] This MI "side band" shifts towards higher wave numbers as the power increases, eventually resulting into thinner and more closely spaced filaments.

In nematic liquid crystals (NLC) a large non-resonant nonlinear response (with an equivalent Kerr coefficient $10^6$-$10^9$ times higher than semiconductors) due to molecular reorientation is coupled to a significant non locality in time and in space.[16-17] When the angle between the average molecular alignment $\hat{n}$ and the field vector $\vec{A}$ is not $\pi/2$ (i.e. Freédericks transition), the reorientation is thresholdless and can sustain the propagation of spatial solitons.[18] Owing to the elastic forces between molecules in NLC, however, the refractive perturbation extends well beyond the driving field, hence the response is highly non local.[19]

Theoretical studies have addressed MI and its features in spatially non local media, [20-21] and the role played by the high non locality of NLC has been recently addressed both theoretically and experimentally.[13, 19, 22] Non locality limits the spectral response, producing a "smooth" transverse perturbation.[12] This corresponds to low-pass filtering in the MI gain, thereby moderating its peak shift towards high wave numbers as the pump power is increased. Spatial modulation instability owing to a spatially incoherent excitation has been studied in weakly non local photorefractive crystals.[6-11]

In this Letter, in order to further understand the interplay between a high non locality and spatial incoherence on the onset of transverse MI, we investigate one-dimensional MI in undoped NLC, comparing its features with the coherent case.

We employed NLC samples as in Ref. [13]: a suitably treated 75µm-thick glass cell contained the nematic uniaxial E7 (indices $n_\perp = 1.53$ and $n_\parallel = 1.77$ and low-frequency dielectric constants $\varepsilon_\perp = 5.1$, $\varepsilon_\parallel = 19.6$) and provided a planar anchorage, while transparent electrodes permitted to voltage bias the NLC and pretilt the molecules at about $\theta_0 = \pi/4$ to eliminate the Freédericks threshold. The beam from an Argon-ion laser was shaped into an ellipse >200µm-wide along the transverse axis Y, according to the geometry sketched in Fig. 1. Light propagation was studied by imaging the light scattered thru the top interface. The spatial coherence of the pump was controlled using a rotating diffuser inserted in the beam path. The diffuser, a plastic disc with scratches in the tangential direction and variable density along the radius, caused spectral broadening in the (Y, Z) propagation plane. The rotation speed of the disc was adjusted to grant a speckle substitution mean time lower than the NLC response time. The degree of spatial incoherence was estimated by a second CCD camera imaging the optical Fourier transform of the input beam and quantified as the ratio between the transverse spectral broadening (standard deviation) with ($\Delta K_Y$) and without ($\Delta K_{Y0}$) diffuser. Figure 2 (left hand side) shows two typical Fourier transforms in the coherent and an incoherent case with $\Delta K_Y = 4.1\, \Delta K_{Y0}$.

Considering a cell much thicker than the beam (across X), the general model governing the reorientation $\Psi$ can be cast in the form: [19]

$$K\frac{d^2\Psi}{dY^2} - \Delta\varepsilon_{RF} E_{RF}^2(\theta_0)\frac{\sin(2\theta_0)}{2\theta_0}(1 - 2\theta_0 \cot(2\theta_0))\Psi + \frac{\varepsilon_0 n_a^2 \langle |A|^2 \rangle}{4}\sin(2\theta_0) = 0 \qquad (1)$$

where $<|A|^2>$ is the time- (or ensemble-) average of the intensity distribution, K the NLC elastic (Frank) constant in the single value approximation (K~$10^{-11}$ N), [17] $\Delta\varepsilon_{RF} = \varepsilon_0(\varepsilon_\| - \varepsilon_\perp)$, $n_a^2 = n_\|^2 - n_\perp^2$, and $E_{RF}$ the electric bias defining $\theta_0$. This model concerns bulk propagation in nematic liquid crystals, in the same experimental geometry of Ref. [13]. The quantitative agreement between theory and data in the coherent case proves that boundary conditions, (e.g. finite sample thickness) play a negligible role.

The theoretical analysis of incoherent MI of Eq.(1) and the paraxial evolution equation for A [13] can be carried out based on the Wigner-Moyal equation with the Klimontovich statistic average.[23] If $\rho_0(p)$ is the Wigner transform describing the noise distribution of the pump (assumed well approximated by a noisy plane wave), and $\rho_1(Y,Z,p)$ its correction due to MI, it follows from (1) that

$$\Psi_1 = \Im_Y^{-1}\left\{\frac{C_1}{K \cdot K_Y^2 + C_2}\int_{-\infty}^{\infty}\Im_Y\{\rho_1(Y,Z,p)\}dp\right\} \quad (2)$$

with $\Psi_1$ the tilt perturbation, $C_1 = \varepsilon_0 n_a^2/4\sin(2\theta_0)$, $C_2 = \Delta\varepsilon_{RF}E_{RF}^2(1 - 2\theta_0\cot(2\theta_0))\text{sinc}(2\theta_0)$, and $K_Y$ the transverse wavenumber. $\Im_Y\{\ \}$ and $\Im_Y^{-1}\{\ \}$ are direct and inverse Fourier transform operators, respectively. (2) represents a Lorentzian low-pass filter, with cutoff wavenumber $K_{Y\_Cut} = \sqrt{C_2/K}$.

When the speckle size in the beam is comparable with the NLC transverse relaxation length, MI is expected to be strongly depressed: high pump-wavenumbers tend to induce transverse perturbations with spatial frequencies larger than the response cutoff, thereby filtered out in propagation and simply contributing to the background illumination.

Following the analysis in Ref. [23], it is possible to write the generalized dispersion relation in the presence of a non local response (details will be given elsewhere):

$$1 + \frac{C_1 C_3}{K \cdot K_Y^2 + C_2}\int_{-\infty}^{\infty}\frac{\rho_0(p + K_Y/2) - \rho_0(p - K_Y/2)}{-iG + K_Y p/k_0}dp = 0 \quad (3)$$

with G the growth rate of the MI unstable plane waves evolving with $\exp(GZ)$, $k_0 = 2\pi n/\lambda$ and $C_3 = k_0 n_a^2 \sin(2\theta_0)/n_0$. For a given $\rho_0(p)$, in general the solution of Eq. (3) is attainable only numerically. In the particular case of a Lorentzian distribution, with linewidth $p_0$, it can be written in closed form ($A_0$ is the peak amplitude of the pump)

$$\frac{k_0 G}{K_Y} = \sqrt{\frac{k_0 C_1 C_3 A_0^2}{K \cdot K_Y^2 + C_2} - \frac{K_Y^2}{4}} - p_0 \qquad (4)$$

Eq. (4) reproduces the coherent-light gain-profile when $p_0 \to 0$ [13] and the result of Ref. [23] in the local limit $K \to 0$. It also implies, as anticipated, that both non-locality and incoherence (measured by the linewidth $p_0$) act to suppress MI.

The comparison shown in Fig. 2 (center and right panels) is in agreement with the predicted trend: as a substantial degree of incoherence is introduced in an otherwise comparable excitation in power, size and average distribution, the insurgence of an MI pattern is severely reduced in visibility over the same propagation distance. Fig. 3 shows a comparison between acquired MI patterns in Z=0.2mm for various degrees of incoherence and excitations. At low power (P=40mW) the beam has an average Gaussian distribution and propagates substantially unmodified in shape. At higher powers (from left to right) MI becomes progressively more visible: a transverse nearly-periodic intensity pattern emerges and eventually breaks up into filaments (P=300mW and $\Delta K_Y = \Delta K_{Y0}$), unless significant incoherence is introduced (top to bottom).

Having defined the MI gain as the ratio between high- and low-power spectra of transverse profiles, Fig. 4 shows that MI is always present, with a general trend in agreement with incoherent-MI in local media: the gain side-band becomes smaller and its peak moves towards lower wavenumbers as the incoherence increases. MI appeared to become unappreciable at lower and lower powers or higher and higher degrees of incoherence, until in our samples the transverse perturbation could no longer be separated from the noisy background.

A direct comparison between Eq. (4) or the more general (3) and experimental data (as in the coherent case, see Ref. [13]) cannot be readily performed from such preliminary results. In the incoherent case, in fact, the shape of the noise spectrum plays a major role and, in general, is not Lorentzian (as assumed in (4)). Moreover, the mutual interplay of incoherence, speckle-size and finite numerical aperture needs be ascertained in much greater detail for a meaningful quantitative assessment.

In conclusions, we have systematically investigated one-dimensional transverse modulation instability in nematic liquid crystals under incoherent excitation. The Wigner transform approach enables a comprehensive model of incoherent MI in highly non local nonlinear media and supports the physical intuition of the interplay between non locality and incoherence. Via the non locality inherent to NLC, the degree of beam spatial incoherence clearly translates into a reduced MI gain with peak shifted to lower wave-numbers. Liquid crystals appear to candidate for optical processing devices in which an adjustable dependence on input (in)coherence can be tailored to a specific functional task.

**Acknowledgements**   We thank C. Umeton and A. De Luca (LICRYL, Rende, Italy) for making the samples available.

**Figure Captions**

Figure 1

Sketch of the NLC cell used in the experiments. The NLC thickness across X was 75µm. An input interface prevents meniscus formation and undesired beam de-polarization. The input beam is shaped into an ellipse with waist $W_Y>200$µm along Y and $W_Y>10W_X$.

Figure 2

Left: two optical transverse input spectra representative of the coherent (top: $\Delta K_Y=\Delta K_{Y0}$) and incoherent (bottom: $\Delta K_Y=4.1\Delta K_{Y0}$) cases. Center: images of the optical intensity propagation in the (Y, Z) plane. Right: transverse spatial profiles in Z=0.2mm.

Figure 3

Transverse beam profiles acquired in 200µm, arranged versus input power (40, 100, 200, 300 mW) and degree of incoherence $\Delta K_Y/\Delta K_{Y0}$ (1, 2.6, 4.1).

Figure 4

MI spectral gain corresponding to the spatial profiles in Fig. 3.

Figure 1

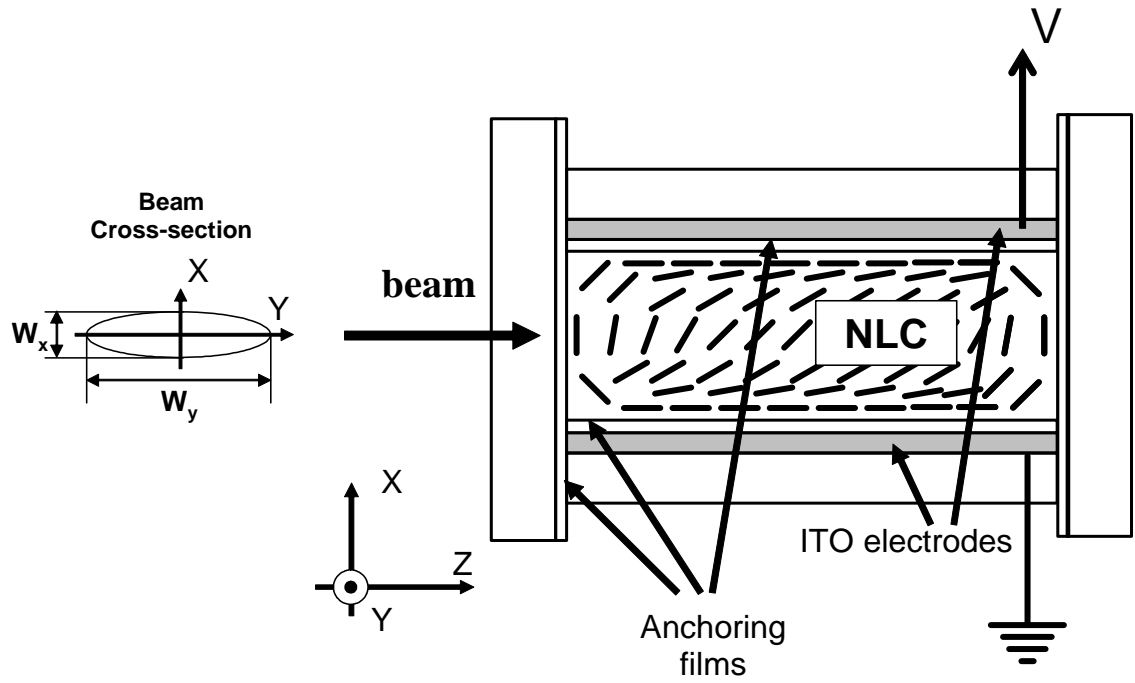

Figure2

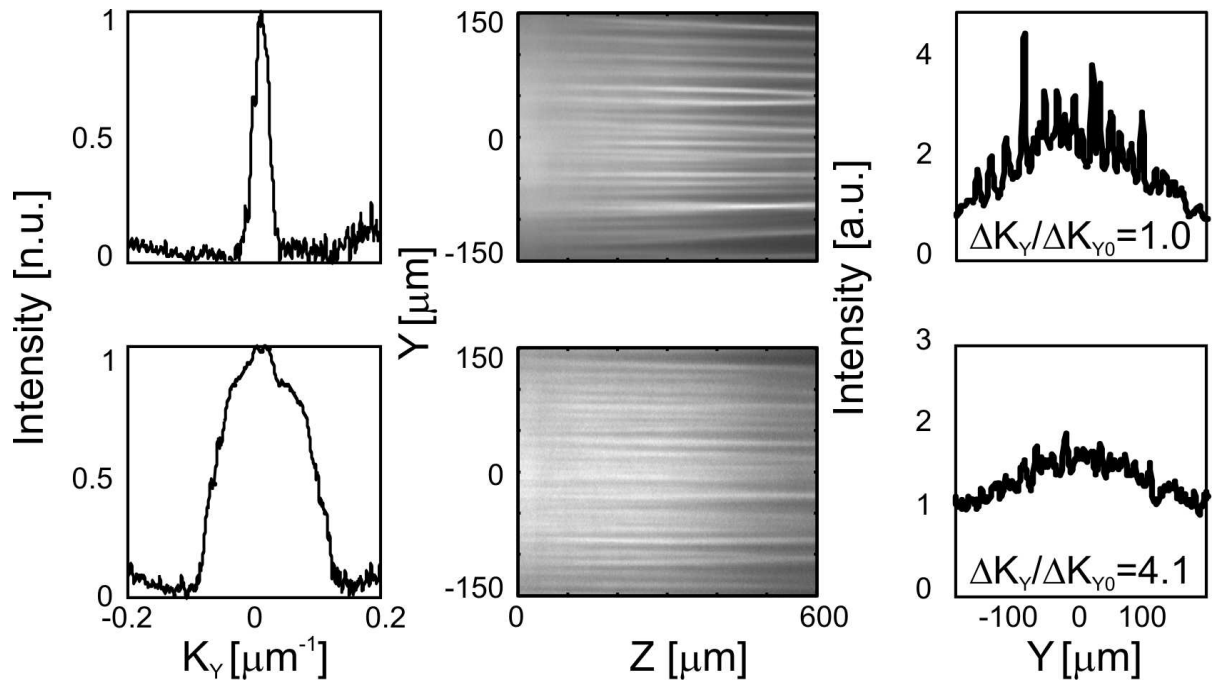

Figure 3

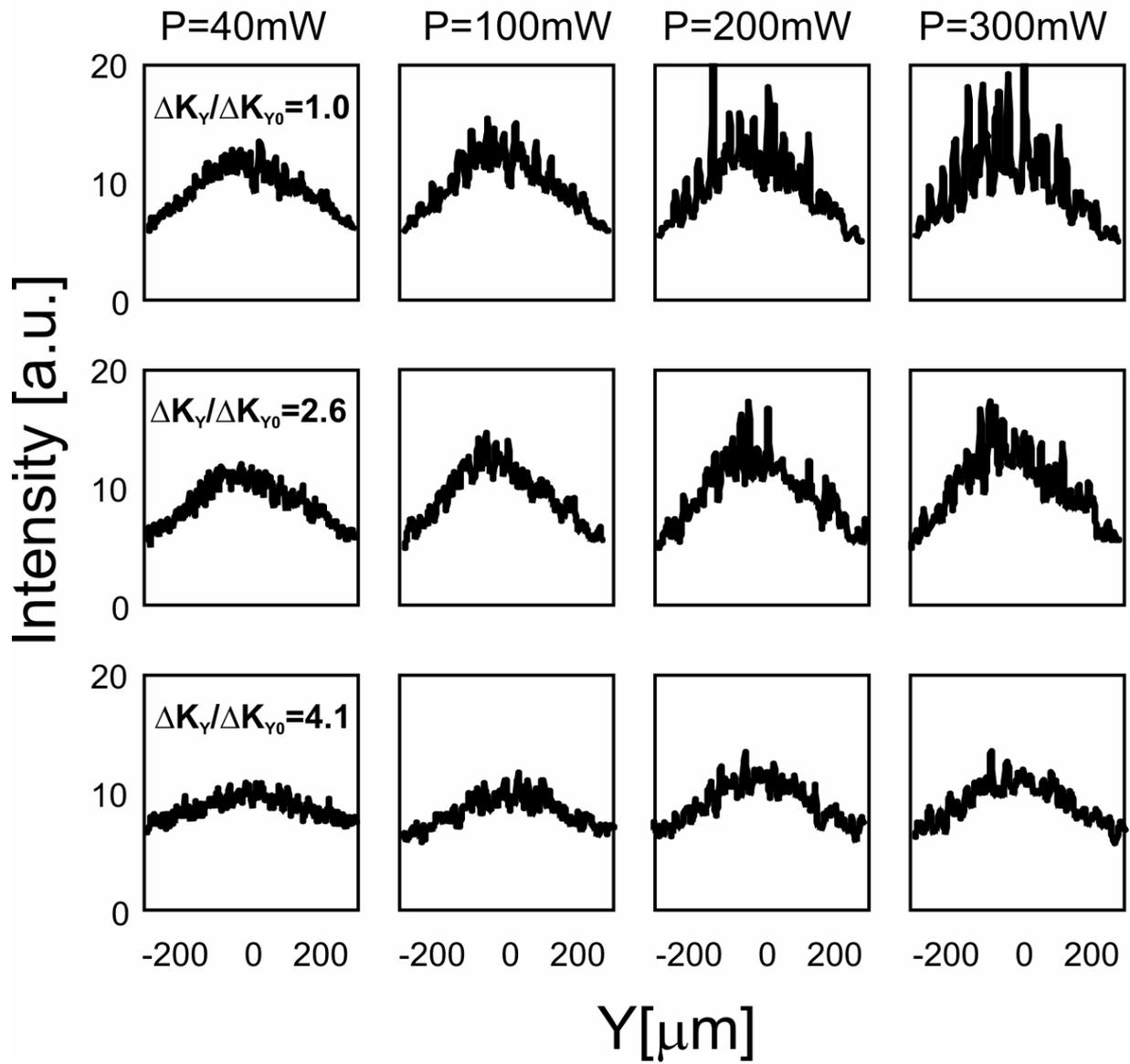

Figura 4

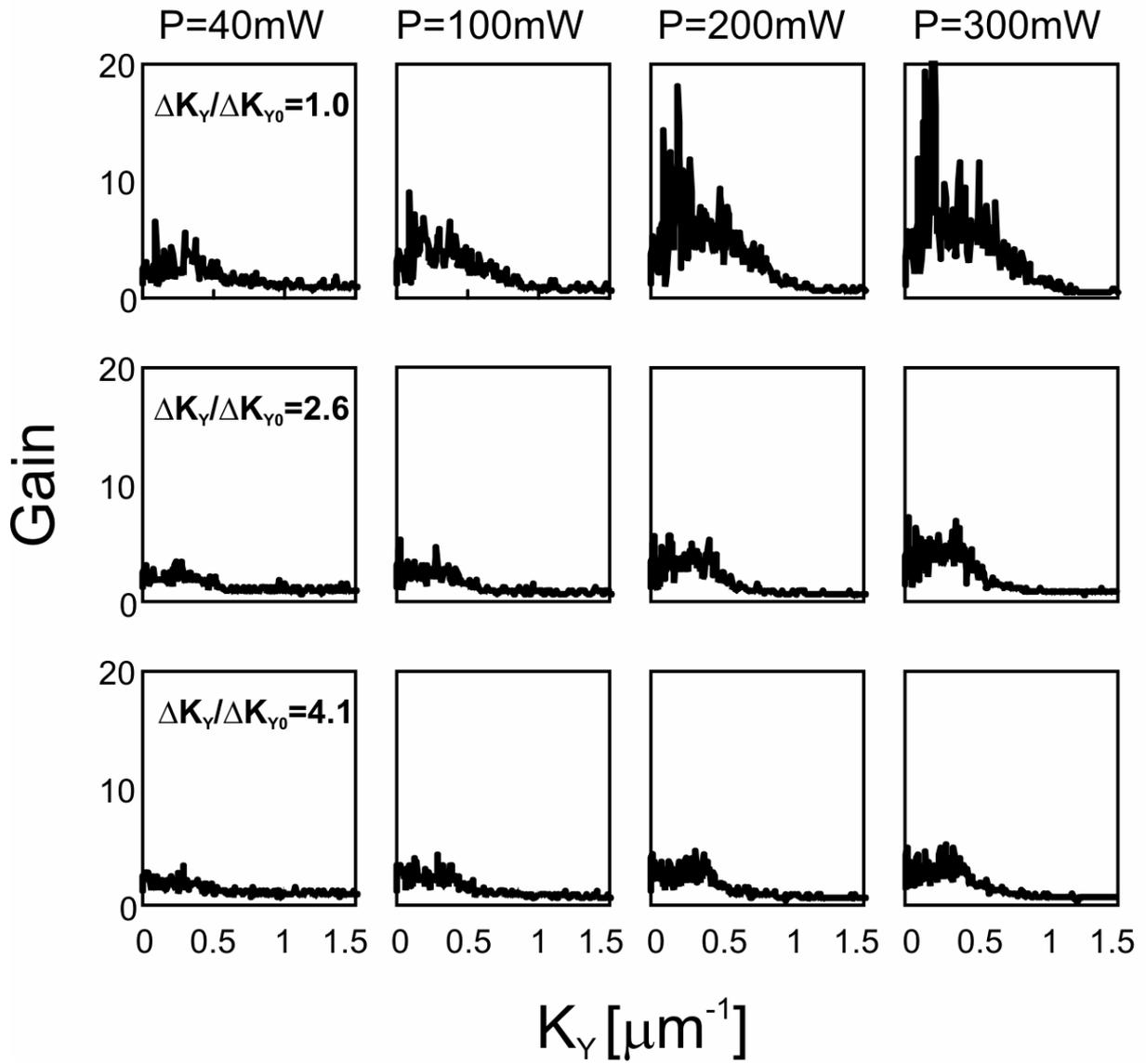